\documentclass[
aps,pra,
reprint,
superscriptaddress,
amsmath,
amssymb
]{revtex4-1}

\usepackage[utf8]{inputenc}
\usepackage{times}
\usepackage{microtype}
\usepackage{graphicx}
\usepackage{upgreek}
\usepackage{bm}
\usepackage{hyperref}
\usepackage{xcolor}
\usepackage[normalem]{ulem}
\usepackage{placeins}
\usepackage{amsmath}
\usepackage{amsthm}
\usepackage{dcolumn}
\usepackage{comment}

\usepackage{filecontents}
\usepackage[T1]{fontenc}
\usepackage{csquotes}
\usepackage[german,ngerman,english]{babel}
\usepackage{centernot}
\usepackage{mathtools}
\usepackage{ifthen}
\usepackage{tikz}
\usetikzlibrary{calc}
\usetikzlibrary{decorations.pathreplacing}
\usetikzlibrary{decorations}
\usetikzlibrary{positioning}
\usepackage{pgfplots}
\usepackage{xxcolor}
\definecolor{structure}{rgb}{0.23,0.4,0.7}
\usepackage[normalem]{ulem}

\newtheorem{lemma}{Lemma}
\newtheorem{theorem}{Theorem}
\newtheorem{corollary}{Corollary}
\newtheorem{definition}{Definition}

\newtheorem{observation}{Observation}

\newsavebox{\blocksavebox}

\definecolor{niceblue}{rgb}{0.33,0.5,0.8}

\newcommand{\refsub}[2]{\hyperref[#1]{\ref*{#1}#2}}

\renewcommand{\max}{\mathchoice{\operatorname*{max}}{\operatorname*{max}}{\mathrm{max}}{\mathrm{max}}} 

\newcommand{\norm}[2][]{
  \ifthenelse{\equal{#1}{}}
    {\left\| {#2} \right\|}
    {\ifthenelse{\equal{#1}{uinv}}
      {\left\vert\kern-0.25ex\left\vert\kern-0.25ex\left\vert {#2} \right\vert\kern-0.25ex\right\vert\kern-0.25ex\right\vert}
      {\left\| {#2} \right\|_{#1}}
    }
}

\newcommand{\taverage}[2][]{
  \ifthenelse{\equal{#1}{}}
  {\overline{#2}}
  {\overline{#2}^{#1}}
}

\newcommand{\tracedistance}[3][]{
  \ifthenelse{\equal{#2}{}}
  {\ifthenelse{\equal{#3}{}}
    {\mathcal{D}_{#1}}{}
  }{
    \ifthenelse{\equal{#1}{}}
    {\mathchoice{\operatorname{\mathcal{D}}\left(#2,#3\right)}{\operatorname{\mathcal{D}}(#2,#3)}{\operatorname{\mathcal{D}}(#2,#3)}{\operatorname{\mathcal{D}}(#2,#3)}}
    {\mathchoice{\operatorname{\mathcal{D}}_{#1}\left(#2,#3\right)}{\operatorname{\mathcal{D}}_{#1}(#2,#3)}{\operatorname{\mathcal{D}}_{#1}(#2,#3)}{\operatorname{\mathcal{D}}_{#1}(#2,#3)}}
  }
}
\newcommand{\fidelity}[3][]{
  \ifthenelse{\equal{#2}{}}
  {\ifthenelse{\equal{#3}{}}
    {\mathcal{F}_{#1}}{}
  }{
    \ifthenelse{\equal{#1}{}}
    {\mathchoice{\operatorname{\mathcal{F}}\left(#2,#3\right)}{\operatorname{\mathcal{F}}(#2,#3)}{\operatorname{\mathcal{F}}(#2,#3)}{\operatorname{\mathcal{F}}(#2,#3)}}
    {\mathchoice{\operatorname{\mathcal{F}}_{#1}\left(#2,#3\right)}{\operatorname{\mathcal{F}}_{#1}(#2,#3)}{\operatorname{\mathcal{F}}_{#1}(#2,#3)}{\operatorname{\mathcal{F}}_{#1}(#2,#3)}}
  }
}

\newcommand{\Sr}[3][]{
  \ifthenelse{\equal{#1}{}}
    {\operatorname{\mathnormal{S}}(#2\|#3)}
    {\operatorname{\mathnormal{S}}_{#1}(#2\|#3)}
}

\DeclareMathOperator{\1}{\mathbb{I}}



\newcommand{\mb}[1]{\mathbb{#1}}

\newcommand{\R}{\mb{R}}
\newcommand{\C}{\mb{C}} 
\newcommand{\one}{\mb{I}}

\definecolor{jens}{rgb}{0.1,0.5,0.1}
\definecolor{martin}{rgb}{0,0,1.0}

\newcommand{\beq}[0]{\begin{equation}}
\newcommand{\eeq}[0]{\end{equation}}

\newcommand{\hide}[1]{}

\begin{document}

\title{Entangling power and quantum circuit complexity}

\author{J.\ Eisert}
\address{Dahlem Center for Complex Quantum Systems, Freie Universit{\"a}t Berlin, 14195 Berlin, Germany}
\address{Helmholtz-Zentrum Berlin f{\"u}r Materialien und Energie, 14109 Berlin, Germany}

\begin{abstract}
Notions of circuit complexity and cost play a key role in quantum computing and simulation where they capture the (weighted) minimal number of gates that is required to implement a unitary. Similar notions also become increasingly prominent in high energy physics in the study of holography. While notions of entanglement have in general little implications for the quantum circuit complexity and the cost of a unitary, in this work, we discuss a simple such relationship when both the entanglement of a state and the cost of a unitary take small values, building on ideas on how values of entangling power of quantum gates add up. This bound implies that if entanglement entropies grow linearly in time, so does the cost. The implications are two-fold: It provides insights into complexity growth for short times. In the context of quantum simulation, it allows to compare digital and analog quantum simulators. The main technical contribution is a continuous-variable small incremental entangling bound.
\end{abstract}

\maketitle

\subsection*{Introduction}
The circuit complexity of a computation captures the number of elementary steps it minimally takes to determine its outcome. A reading of the famous Church-Turing thesis states that all reasonable models of computation give rise to the same class of ``easy'' problems computable in 
polynomial time, a statement that can presumably also be applied to processes occurring in nature.
Alas, ultimately the world is quantum. Indeed, notions of quantum circuit complexity have long been
considered in quantum information science: They provide a quantitative account on the shortest quantum 
computation that implements a given unitary. Similarly, one can think of the complexity of a quantum
state as the circuit complexity of the quantum circuit preparing it, starting from a given fiducial state.
Such notions play a similarly central role in quantum  as 
classical circuit complexities do in classical computing. 
Seminal work 
\cite{Nielsen:2005mn1,Nielsen:2006mn2,PhysRevA.73.062323,Nielsen2007}
has introduced a geometric picture of circuit complexities, showing that finding the shortest circuit
amounts to identifying the shortest path between two points in a  curved geometry. In fact, this
program has become so successful that the cost associated with a unitary in such a  
geometric picture has itself been identified with a notion of circuit complexity. 


Yet, it was relatively recently that notions of circuit
including those of costs rose to prominence outside the
field of quantum computing \cite{PhysRevD.90.126007,PhysRevLett.116.191301,ChapmanMarrochioMyers,Brown:2017jil,Reynolds:2016rvl,Reynolds:2017lwq,Complexity1,JeffersonMyers,BigComplexity,AaronsonComplexity,ComplexityGrowth}. Again eluding to the physical Church Turing thesis, such an
approach is well motivated: One can think of a quantum state -- say, one that is being
generated by a quantum chaotic Hamiltonian evolution -- being highly complex if the quantum circuit that 
could have prepared it on a quantum computer would have to be long. Since one can 
argue about how many quantum gates one would have needed to emulate a given Hamiltonian time evolution,
such notions also immediately allow to compare the effort in digital and analog quantum simulation
\cite{CiracZollerSimulation}.
The possibly most compelling application of quantum circuit complexity is in the 
 realm of high energy physics in the context of holography
\cite{PhysRevD.90.126007,ChapmanMarrochioMyers,PhysRevLett.116.191301,Brown:2017jil,Reynolds:2016rvl,Reynolds:2017lwq,Complexity1,JeffersonMyers,BigComplexity,AaronsonComplexity,ComplexityGrowth}. 


 \begin{figure}[tb]
\centering
\includegraphics[width=0.8\columnwidth]{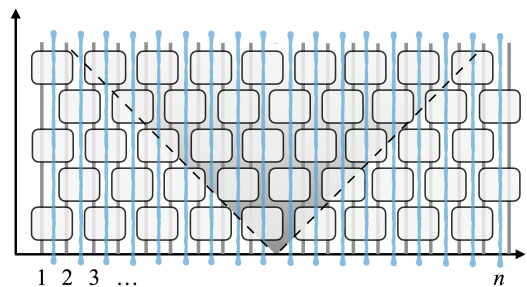}
\caption{A schematic picture relating the cost of a circuit with the entanglement over
cuts  for a system of $n$ constituents. The dark grey triangle represents the Lieb-Robinson
cone \cite{Hastings-CMP-2006,EisertOsborne06,BravyiHastingsVerstraete06} 
that depicts at what rate one expects a linear growth of the entanglement entropy over all cuts
in non-equilibrium dynamics generated by a local Hamiltonian.}
\label{Figure}
\end{figure}

 These thoughts provide fuel for a motivation to actually compute quantum circuit complexities
 and circuit costs. Yet, to
 actually quantitatively determine any variant of these quantities is not obvious. After all, there are 
 many ways to decompose a given unitary into a quantum circuit, with the  
 best known algorithms for decomposing given circuits in Clifford and $T$-gates
 featuring an exponential run-time
 in the circuit size \cite{TCount}, and
 the computation of the complexity requires the optimization over such decompositions. 
 In any decomposition,
 one may expect cancellations of some sort, with the impact of a unitary gate being partially compensated
 by the later action of another, rendering naive
combinatoric arguments involved.
 The geometrically motivated notion of a cost of a quantum circuit substantially lessens the
 technical burden \cite{Nielsen:2005mn1,Nielsen:2006mn2,PhysRevA.73.062323,Nielsen2007}, 
 but it is still not obvious how 
to come up with meaningful lower bounds.
 
 This work provides a compellingly simple lower bound for the cost of a quantum circuit
  that is tight
  for small values of the cost. It has indeed rightfully been argued that complexity is not entanglement
  \cite{AaronsonComplexity}, and neither is the cost of a circuit. No
  quantity based on entanglement can accommodate the presumed linear growth of state complexity
  until a time exponential in the system size \cite{PhysRevD.90.126007}, for obvious reasons. 
  That said, for small values of 
  the circuit cost and entanglement there is a simple connection: One can 
  basically add up -- if properly put together -- 
  potential entangling powers of quantum gates to arrive at tight bounds. 
The bounds presented 
are rooted in notions of entanglement capabilities of quantum gates: 
The argument 
captures the insight that quantum gates that are close
to the identity in operator norm have little capability to create entanglement from product states (which is very easy
to show). They can also add very little entanglement to a given entangled state
(which is less obvious to prove), but this can be grasped in terms of the small incremental entangling property \cite{Marien2016,PhysRevLett.111.170501}, 
and which is here freshly proven for Gaussian continuous-variable systems. As such, the simple bound applies both to spin systems as to Gaussian bosonic
continuous-variable settings, which are specifically important when approximating non-interacting bosonic
quantum fields. Simple as the bound is, it is easily stated and proven (with some of the arguments delegated to the appendix). It can also straightforwardly be applied to important cases of quantum evolutions,
for which quenched Hamiltonian many-body dynamics constitute an example.

\subsection*{Quantum circuit complexity and cost}

The exact circuit complexity basically counts the number of quantum gates 
from a given gate set that is needed to exactly match the given unitary. 
An approximate reading thereof 
merely asks for an approximation in operator norm to a given small error.
Lower bounds of the circuit complexity are provided by the 
\emph{cost
of a given circuit}, which is increasingly commonly seen as a
notion of circuit complexity in its own right
\cite{Nielsen:2006mn2,Nielsen:2005mn1}. For $n$ quantum systems of local dimension $d$ ($d=2$ for spins or qubits), one chooses a collection of 
$2$-local traceless Hamiltonian terms $O_1,\dots, O_J$, normalized in operator norm
as $\|O_j\|=1$ for $j=1,\dots, J$.
We consider both the situation in which $\{O_j\}$ are geometrically
local and the situation where they are merely local in their support. 
For a given $U\in$ SU$(d^n)$, one
regards the unitary as being generated by a path-ordered integral 
\begin{equation}\label{GD}
 U = {\cal P} \exp \left(-i \int_0^1 ds H(s) \right),
\end{equation}
with 
\begin{equation}\label{GD2}
	H(s) = \sum_{j=1}^J y_j (s) O_j,
\end{equation}
where $y_j:[0,1]\rightarrow \R$ are appropriate continuous cost functions. 
This
path ordered integral can in operator norm arbitrarily well approximated by 
 \begin{eqnarray}\label{TO}
	V_N&=&\prod_{k=1}^N \exp \left(-\frac{i}{N}  \sum_{j=1}^J y_j (k/N) O_j   \right)
\end{eqnarray}
in the limit of $N\rightarrow \infty$,
as follows immediately from the definition of the path-ordered integral.
The cost of a unitary $U\in$ SU$(d^n)$ 
can then be defined in such terms  \cite{Nielsen:2005mn1,Nielsen:2006mn2}. 

\begin{definition}[Circuit cost \cite{Nielsen:2005mn1}]
For a given set $\{O_1,\dots, O_J\}$ in the Lie algebra su$(d^n)$ 
of traceless Hermitian matrices normalized as $\|O_j\|=1$ for all $j=1,\dots, J$,
the \emph{cost of a quantum circuit} $U\in $SU$(d^n)$
is the infimum  
\begin{equation}
	C(U):= \inf \int_0^1 \sum_{j=1}^J | y_j(s)| ds
\end{equation}
over all continuous functions $y_j:[0,1]\rightarrow \R$ 
so that Eqs.\ (\ref{GD}, \ref{GD2}) are satisfied.
We call it the geometrically
local circuit cost $C_{\rm g}(U)$ 
if all $\{O_j\}$ are geometrically local.
\end{definition}
That is to say,
the cost of a quantum circuit can be expressed in terms of the limit
\begin{equation}
	 \lim_{N\rightarrow \infty}\frac{1}{N}\sum_{k=1}^N   \sum_{j=1}^J | y_j (k/N) |
\end{equation}
of many time steps.

\subsection*{Potential entangling power} 
In what follows, the notion of a  potential entangling power of a quantum gate provides some useful intuition. 
It captures
the ``coupling strength'' and simply takes into account the fact that quantum 
gates that are close to the identity cannot create 
much entanglement. A somewhat related, but integer-valued, 
notion of entangling power
has been invoked in Ref.\  \cite{Balasubramanian}.

\begin{definition}[Potential entangling power] A unitary $U\in $SU$(d^2)$ has the
potential entangling power
\begin{equation}
	e(U) := \log(d) \,\text{min}\left\{\|H\|:  U=e^{-i H }, H=H^\dagger\right\}.
\end{equation}
\end{definition}
It is indeed perfectly meaningful to refer to this quantity as the  potential entangling power:
If $\rho = \rho_A\otimes \rho_B$, both $\rho_A$ and $\rho_B$ being pure and supported on $\C^d$ each, then 
the resulting degree of entanglement 
	$S({\rm tr}_B (U\rho U^\dagger))$
as quantified in terms of the von-Neumann \emph{entanglement entropy} over the cut $A:B$
is expected to be small if $e(U)$ is small, and converging to zero for $e(U)\rightarrow0$.
Notions of entangling powers of quantum gates have long been connected to coupling strengths
of interactions \cite{PhysRevA.62.052317,PhysRevA.64.032302,PhysRevLett.86.544,PhysRevA.62.030301}.
As is well-known, notions of entangling power of unitary gates are altered depending on whether or not
auxiliary quantum systems are allowed for: The swap gate obviously has no 
entangling power, if no auxiliary systems made use of,
while it has $2 \log_2(d)$ when auxiliary systems are included. It is less obvious to see how much entanglement can be generated, however, if one initially already encounters an intricate entangled state and the
unitary acts only on a small subsystem of the total system. The question of how much entanglement can
be generated in this fashion has  been largely
settled in Ref.\ \cite{Marien2016}, however, which we can make use of here.

\subsection*{Entanglement bounds circuit costs} 
Such notions of potential entanglement power can be related to tight bounds of
circuit costs. In what follows, we denote for a pure state $\rho$ defined on 
a spatially one-dimensional system of $n$ constituents for $s\in\{1,\dots, n-1\}$ with 
\begin{equation}
 E(\rho: s) := S({\rm tr}_B (\rho))
 \end{equation}
 being 
the \emph{entanglement entropy} 
over the cut $A= \{1,\dots, s\}$ and
 $B= \{s+1,\dots, n\}$. 
\begin{observation}[Entanglement lower bounds the cost]
The geometrically local circuit cost of a  $U\in $SU$(d^n)$ 
 is lower bounded by
\begin{equation}\label{RHS}
 C_{\rm g}(U) \geq \frac{1}{c \log(d)}\sum_{s=1}^{n-1} 
 E(U|\phi\rangle\langle \phi  | U^\dagger: s) ,
\end{equation}
for an absolute constant $c>0$, where $|\phi\rangle\in (\C^d)^{\otimes n}$ is a product
state vector. 
For the cost, one finds
\begin{equation}\label{RHS}
 C (U) \geq  \frac{1}{c \log(d)}\max_s  
 E(U|\phi\rangle\langle \phi  | U^\dagger: s) .
\end{equation}
\end{observation}
The \emph{potential 
cancellation} of gates in notions of complexity is faithfully captured in this bound: If most gates in a circuit commute, they will give rise to a lower circuit cost,
 but at the same time also to a smaller entanglement. So even if the bound is simple indeed, it does capture a 
key feature of the relationship of the circuit cost to notions of entanglement.

\begin{proof}
%
The proof of this observation is straightforward, acknowledging the results of Ref.\ \cite{Marien2016}. 
We start by decomposing the circuit in a convenient manner. Making use of a Trotter decomposition, we find
that $U$ can 
in operator norm  $\| U-W_N\|$ be
arbitrarily well approximated as a product
 \begin{eqnarray}
W_N &:= &\prod_{k=1}^N V_k
\end{eqnarray}
with  each term being given by
 \begin{eqnarray}
V_k & :=&  \exp \left(-\frac{i}{N}  \sum_{j=1}^J y_j (k/N) O_j   \right)\\
&=& \lim_{m\rightarrow\infty}
\left(
V_{k,1}^{1/m}
\dots
V_{k,J}^{1/m}
\right)^m\nonumber
\end{eqnarray}
where
 \begin{eqnarray}
V_{k,j}:= 
\exp \left(-\frac{i}{N}   y_j (k/N) O_j  \right).
\end{eqnarray}
Building upon this, let 
\begin{eqnarray}
|\psi_l\rangle := \prod_{k=1}^l V_k |\psi\rangle
\end{eqnarray}
be the state vector after $l\in \{1,\dots, N\}$
temporal layers, 
with $|\psi_0\rangle:= |\phi\rangle$. Then, for $l=1,\dots, N$, using the time integrated instance
of Lemma \ref{V},
one finds that the entanglement growth over the cut
$A=\{1,\dots, s\}$ and $B=\{s+1,\dots, n\}$
in each step
can at most be
\begin{eqnarray}
	 E(|\psi_{l}\rangle\langle \psi_{l}|&:&s)
	 - 
	E(|\psi_{l-1}\rangle\langle \psi_{l-1}|: s)\nonumber\\
	&=& E(  V_l  |\psi_{l-1}\rangle\langle \psi_{l-1}|  V_l^\dagger  :s)
	 - 
	E( |\psi_{l-1}\rangle\langle \psi_{l-1} |
	: s)\nonumber\\
	&\leq & \frac{m c}{N} \sum_{j=1}^J  \frac{1}{m} y_j (l/N) \|O_j\| \log(d)
\end{eqnarray}
which gives	
	\begin{eqnarray}
	 E(|\psi_{l}\rangle\langle \psi_{l}|&:&s)
	 - 
	E(|\psi_{l-1}\rangle\langle \psi_{l-1}|: s)\nonumber\\
	&\leq & \frac{c \log(d)}{N} \sum_{j=1}^J   |y_j (l/N)| .
\end{eqnarray}
Iterating this expression, one finds
\begin{eqnarray}
	&& E(U |\phi\rangle\langle \phi| U^\dagger :s)
	 - 
	E(|\phi\rangle\langle \phi|: s)\nonumber\\
	&\leq & \frac{c \log(d)}{N}  \sum_{k=1}^N\sum_{j=1}^J   |y_j (l/N)| .
\end{eqnarray}
Acknowledging that the right hand side approximates the circuit cost $C(U)$ 
arbitrarily well, find finds the statement of Observation 1, by applying the argument to the cut 
$A=\{1,\dots, s\}$ and $B=\{s+1,\dots, n\}$ providing the tightest bound. For the 
geometrically local circuit cost $C_{\rm g}(U)$, 
the argument can be applied to each such cut,
leading to the statement of Observation 1.
\end{proof}
In the above statement, the following statement from Ref.\ \cite{Marien2016} has been made use of.

\begin{lemma}[Small incremental entanglement \cite{Marien2016}]\label{V}
For a pure state $\rho$ and a Hamiltonian $h$ supported on a $d\times d$-dimensional
subspace acting over the cut $\{1,\dots, s\}$ and $\{s+1,\dots, n\}$,
the entangling rate defined as
 \begin{equation}
 	\Gamma(h,\rho) := \left. \frac{d}{dt} 
	E\left(e^{-ith} \rho e^{ith}: s
	\right)
	\right|_{t=0}
 \end{equation}
 is upper bounded by 
\begin{equation}
	\Gamma(h,\rho) \leq c \log(d) \|h\|.
\end{equation}
\end{lemma}
The constant presented in the proof is $c=22$, but numerical evidence
is shown that rather $c=2$ actually provides a tight bound.
Interpreted in terms of the above notion of an potential entangling power of a unitary
$X\in U(d^2)$ acting on two constituents connecting the subsystems over the cut, 
one can argue that 
\begin{equation}
	| E(X \rho X^\dagger : s) - E(  \rho   : s)|\leq c e(X) ,
\end{equation}
so that up to an absolute constant, the maximum increase of entanglement is indeed nothing but the 
potential entangling power: In each application, a quantum gate with a certain potential entangling power can increase the value of entanglement only to some extent, no matter how entangled the initial state has been. From the above Trotter decomposition it also follows that the circuit cost is nothing but the
weighted quantum circuit complexity, weighted by the potential entangling power of each
quantum gate.

\begin{corollary}[Weighted quantum circuit complexity] For a given $U\in$SU$(2^n)$, the 
infimum of the sum of weights $e(U_j)$ of a circuit consisting of quantum gates $\{U_j\}$
 generated by $\{O_j\}$ is given by $C(U)$.
\end{corollary}

\subsection*{Gaussian circuit cost}
In fact, there is a small incremental entanglement bound as well as a 
harmonic equivalent of the above relationship between entanglement and quantum circuit
cost for Gaussian bosonic settings \cite{Complexity1,BigComplexity},
including ones motivated by evolutions of non-interacting bosonic quantum fields. For such bosonic
systems, characterized by canonical coordinates
$R= (x_1,p_1, x_2,p_2,\dots, x_n,p_n)$, the appendix presents the proof of the following
small incremental entanglement statement for such continuous-variable systems.

  \begin{theorem}[Gaussian small incremental entanglement] 
For a pure Gaussian state $\rho$ and a Hamiltonian $H= R h R^T$ supported on one of the modes each
of $A= \{1,\dots, s\}$ and $B=\{s+1,\dots, n\}$,
the entangling rate defined as
 \begin{equation}
 	\Gamma(h,\rho) := \left. \frac{d}{dt} 
	E\left(e^{-itH} \rho e^{itH}: s
	\right)
	\right|_{t=0}
 \end{equation}
 is upper bounded by 
\begin{equation}
	\Gamma(h,\rho) \leq \|h \|  f(\| \gamma(0)\| ),
\end{equation}
where $f:[1,\infty)\rightarrow \R$ is a monotone increasing function.
\end{theorem}
Interestingly, it is not the operator norm of the Hamiltonian as such (which would make little sense anyway and would not be finite) but that of the kernel matrix when expressed as a polynomial in canonical coordinates that features in this small incremental
entanglement statement. In the same way as above, and elaborated upon in the appendix,
we can conclude the following.

\begin{observation}[Gaussian entanglement lower bounds Gaussian circuit cost]
The geometrically local Gaussian quantum circuit cost of a bosonic Gaussian unitary 
$U$ that prepares $U |\phi\rangle$
from the product  state vector $|\phi\rangle$ associated with the covariance matrix 
$\gamma(0)$
is lower bounded by
\begin{equation}\label{GRHS}
 G_{\rm g}\geq \frac{1}{f(\|\gamma(0)\|)}\sum_{s=1}^{n-1} 
 E(U |\phi\rangle\langle \phi| U^\dagger: s) .
\end{equation}
For the Gaussian quantum circuit cost one finds
\begin{equation}\label{GRHS}
 G \geq  \frac{1}{f(\|\gamma(0)\|)} \max_s  
 E(U |\phi\rangle\langle \phi| U^\dagger: s) .
\end{equation}
\end{observation}
Making use of these statements, one can infer about non-interacting bosonic theories in largely the same way as for
spin systems, despite unbounded operators featuring in the problem.

\subsection*{Quenched quantum many-body systems}
Simple as the above bounds are, they provide tight and relevant bounds to circuit costs
and complexities for small times in a number of settings. An interesting insight along these lines of thought
is the point that whenever a quantum many-body system 
undergoing \emph{non-equilibrium dynamics} leads 
to a linear increase in the entanglement entropy over suitable cuts, so does the quantum state
complexity. This is in particular true for quenched quantum many-body systems, for which the linear growth of entanglement entropies is generic \cite{1408.5148,PolkovnikovReview,christian_review}. 
In fact, both upper \cite{EisertOsborne06,BravyiHastingsVerstraete06} and lower bounds
\cite{Schuch_MPS} for the entanglement entropy as a function of time have readily been 
established. That is to say, 
whenever the right hand side of Eq.\ (\ref{RHS}) grows linearly in time, so does the left hand side, as an immediate corollary (see Fig.\ \ref{Figure}). 
We state this explicitly for the Ising Hamiltonian, but it should be clear that the same behaviour
is expected for any local Hamiltonian (not featuring disorder).

\begin{observation}[Growth of circuit cost in dynamics] For any time $T>0$ there exists a system size $n$ 
for a
translationally invariant Ising Hamiltonian such that 
the unitary dynamics $e^{-itH}$ applied to a product state vector $|\phi\rangle$ leads to 
$C(e^{-iH t})> \delta t$ for an absolute constant $\delta>0$, 
for all times $t\in [0,T]$.
\end{observation}
The upper bound in time $T$ is merely accommodating the possibility of having a finite system of finitely many degrees of freedom $n$, for which at some point, the respective entanglement entropies will no longer 
grow in time (rendering the bound then uninteresting). 
The result stated here is a corollary of Observation 1, together with the results of Ref.\ \cite{SchuchQuench}. Since the model is translationally invariant, any cut serves 
to show the linear growth of 
the quantum state complexity in time. For the geometrically local circuit cost, one also finds a
growth linear in time, but now the largest value of $C_{\rm g}(e^{-iH T})$ attained at intermediate times 
scales as $\Theta(n^2)$ in the system size $n$, instead of the essentially 
linear scaling $\Theta(n)$ in case of the quantity 
$C(e^{-iH T})$. 

\subsection*{Summary and outlook}
In this work, we have carefully and quantitatively 
revisited the connection between entanglement and notions of circuit cost and 
complexity. While there is in general no tight connection between these quantities, for small
values, there actually is, as this work shows: Indeed, one arrives at compellingly simple bounds. The usefulness 
of such bounds is manifest. One can argue, for example, how deep a weighted 
quantum circuit has to be to give rise
to a given entanglement pattern in a desired final state; this is true at least for pure states, but it 
seems perfectly conceivable that similar techniques can be established for mixed quantum states. Also, it helps assessing the power and capabilities
of analog quantum simulators \cite{CiracZollerSimulation}. 
Using such tools, one can argue that a digital quantum simulator would have 
required a precisely defined computational effort to produce the same results as a 
given analog quantum simulator. In this sense, it makes the computational effort of
digital and analog quantum simulators comparable. It is the hope that this simple bound provides 
a useful and versatile tool in various studies of this kind.

\subsection*{Acknowledgements} 
I would like to warmly thank 
V.\ Balasubramanian,
B.\ Chairo,
S.\ Chapman, 
L.\ Hackl, 
R.\ C.\ Myers, 
and specifically M.\ Heller for discussions.
This work has been supported by the DFG (CRC 183, project A03 and B01, FOR 2724, EI 519/14-1, and
EI 519/15-1), the BMBF (DAQC), and the FQXi. It has also received funding from 
the European Union's Horizon 2020 research and innovation programme 
under grant agreement No. 817482 (PASQuanS).

\section*{Appendix}
\subsection*{Preliminaries}
Consider a quantum system of $n$ bosonic modes, equipped with the canonical coordinates
$R= (x_1,p_1, x_2,p_2,\dots, x_n,p_n)$ reflecting positions and momenta. For a
quantum system comprising of $n$ modes, 
the canonical commutation relations give
rise to a
\emph{symplectic form}
\begin{equation}\label{symp}
	\sigma = \bigoplus_{j=1}^n\left[
	\begin{array}{cc}
	0 & 1\\
	-1 & 0
	\end{array}
	\right].
\end{equation}
We consider \emph{Gaussian states} $\rho$ 
\cite{Continuous,GaussianQuantumInfo} 
with vanishing first moments (which can be assumed to be the case
without loss of generality in the context considered) and second moments that can be captured in 
the \emph{covariance matrix} $\gamma\in \R^{2n\times 2n}$ with entries
\begin{equation}
	\gamma_{j,k} = {\rm tr}(\rho (R_j R_k  + R_k R_j)).
\end{equation}
Covariance matrices always satisfy the \emph{Heisenberg uncertainty principle}
$\gamma + i \sigma\geq 0$: It takes a moment of thought that this is nothing but
the standard Heisenberg uncertainty principle written in a way that is manifestly
invariant under symplectic transformations that preserve the symplectic form $\sigma$.

In the endeavour of bounding quantum circuit complexities,
the above quantification in terms of operator norms no longer makes sense. 
However, when assessing notions of circuit cost and complexity, similar bounds 
can still be derived when appropriately evaluated for Hamiltonian terms. To this goal,
let $\{O_j:j=1,\dots , J\}$ be a collection of operators -- with no assumption on the
cardinality $J$ of the set being made -- each of which being of the form
\begin{equation}\label{Each}
	O_j = (X_k,P_k,X_l, P_l) h_j (X_k,P_k,X_l, P_l)^T,
\end{equation}
where $k\in\{1,\dots, s\}$ and
$l\in\{s+1,\dots, n\}$ are labels referring to modes
in $A$ and $B$, respectively. 
 The fact that the indices $k,l$ are from the set of modes reflects the feature that the
operators are $2$-local, but as before, geometric locality may or may not be 
assumed. The
matrices $h_j\in \R^{2\times 2} $ satisfy
\begin{equation}
	h_j=h_j^T,\, \|h_j\|=1. 
\end{equation}	
That is to say, it is no longer the operators
$O_j$ as such that have unit operator norm, but rather the kernels of quadratic operators in the 
canonical coordinates. Equipped with this preparation, we can again
think of Hamiltonians 
\begin{equation}
	H(s) = \sum_{j=1}^J y_j (s) O_j,
\end{equation}
with as before $y_j:[0,1]\rightarrow\R$ being arbitrary continuous functions. Just in the same way considered
above,
the Gaussian circuit cost of a Gaussian unitary $U$ becomes
\begin{equation}
	 G(U) =\lim_{N\rightarrow \infty}\frac{1}{N}\sum_{k=1}^N   \sum_{j=1}^J | y_j (k/N) |,
\end{equation}
or analogously $G_{\rm g}(U) $, depending on whether the generators have been chosen merely local or additionally
geometrically local.
We are now in the position to prove Theorem 1.

\subsection*{Proof of Theorem 1}

\begin{proof}
In what follows, we prove the Gaussian bosonic small incremental entanglement statement summarized
in Theorem 1 of the main text. 
As before, 
 we consider a bi-partite system consisting of parts $A=\{1,\dots, s\}$ 
 and $B=\{s+1,\dots, n\}$ 
 with a two-local Hamiltonian 
supported on modes labeled $k\in A$ and $l\in B$, with the rest of the systems
 \begin{equation} 
 a=\{1,\dots, s \}\backslash 
 \{k\}
 \end{equation} 
 and $b=\{s+1,\dots, n\}\backslash\{l\}$ 
 acting as auxiliary systems. The entanglement rate 
 for a given state $\rho$ is then 
 \begin{equation}
 	\Gamma(H,\rho) := \left. \frac{d}{dt} 
	E(\rho(t):s)
	\right|_{t=0}
 \end{equation}
 where
 \begin{equation}
 	\rho (t) =  
	e^{-itH}
	\rho
	e^{itH}.
 \end{equation}
 On an abstract level,
 this entangling rate has been shown in Ref.\ \cite{Marien2016} to be given by
 \begin{equation}	
	\Gamma(H,\rho) = - i{\rm tr}( ( \1_a\otimes H ) [\rho_{A,\{l\}}, \log ({\rho_{A}})
 \otimes \1_{\{l\}}]).
 \end{equation}
The 
Hamiltonian 
\begin{equation}
	H = R h R^T,
\end{equation}
which is a quadratic polynomial in the bosonic operators $R$,
acts non-trivially on two modes in $A$ and $B$ each, 
so that $h\in \R^{2n\times 2 n}$ features non-vanishing
elements only within a $4\times 4$ block reflecting pairs of canonical coordinates
of two modes and has ${\rm rank}(h)=4$ (more common is a coupling in position only, so that
the respective block in the momentum sector is proportional to $\one$). 
The generalization to
$k$-local Hamiltonians with Hamiltonian kernels of rank $2k$ is immediate
and only omitted for notational convenience.

The subsequent steps can be proven entirely on the level of level of second  moments.
The covariance matrix $\gamma$ of a pure Gaussian state $\rho$ 
takes the form
\begin{equation}
\gamma =\left[
\begin{array}{cc}
\gamma_{A} & \Xi\\
\Xi^T & \gamma_{B}
\end{array}
\right],
\end{equation}
where the principal sub-matrices $\gamma_{A} $ and $\gamma_{B}$ reflect the 
reduced quantum states of the
sub-systems labeled $A$ and $B$, respectively.
The entanglement entropy of $\rho$ with respect to the split of $A$
versus $B$ \cite{AreaReview}
can be computed as
\begin{eqnarray}
E(\rho:s)
&=&\sum_{k=1}^{s} 
    \biggl[ \left(\frac{\sigma_{k} +1} 2 \right)\log \left(\frac{\sigma_{k} +1} 2 \right)\nonumber\\
    &-& \left(\frac{\sigma_{k} -1} 2 \right)\log \left(\frac{\sigma_{k} -1} 2 \right) \biggr] ,
\end{eqnarray}
where $\sigma_1,\dots, \sigma_s\geq 1$ are the \emph{symplectic eigenvalues} of the $s$ modes of
$\gamma_{A}$. These symplectic eigenvalues are
the positive square roots of the eigenvalues of
$-(\gamma_{A} \sigma_{A} )^2$, where
\begin{equation}
	\sigma_{A} := \bigoplus_{j=1}^s
	\left[
	\begin{array}{cc}
	0 & 1\\
	-1 & 0\\
	\end{array}
	\right]
\end{equation}	
is the symplectic
form of the $s$ modes constituting $A$, so that the entanglement entropy is found to be
\begin{eqnarray}
S_{A}(\rho) &=&{\rm tr}
    \biggl[ \left(\frac{M_{A}^{1/2} +\1} 2 \right)\log \left(\frac{M_{A}^{1/2} +\1 } 2 \right)\nonumber\\
    &-& \left(\frac{M_{A}^{1/2}-\1} 2 \right)\log \left(\frac{M_{A}^{1/2} -\1} 2 \right) \biggr] ,
\end{eqnarray}
where $M_{A}$ is the Hermitian $s\times s$-matrix defined as
\begin{eqnarray}
	M_{A}:= \gamma_{A}^{1/2} (i\sigma_{A})  \gamma_{A}  (i\sigma_{A})  
	\gamma_{A}^{1/2}
\end{eqnarray}
in terms of matrix square roots of covariance matrices.
In the same way, $M_{A}(t)$ can be defined for all times $t\geq 0$, derived from the
 second moments of 
$\rho_{A}(t)$ given by 
\begin{equation}
	\gamma_{A}(t) = \left.e^{- \sigma h t}\gamma e^{ \sigma h t}\right|_{A},
\end{equation}
$\sigma$ denoting the symplectic form of the entire system involving all $n$ modes
as defined in Eq.\ (\ref{symp}).
Prepared in this fashion, we can turn to actually upper bounding the incremental
entanglement rate. An explicit calculation shows that 
\begin{eqnarray}
	\Gamma(H,\rho) &=&\frac{1}{2}
	{\rm tr}
    \biggl[ \frac{d}{dt}\left. M_{A}^{1/2} (t)   \right|_{t=0}\\
    &\times & \left(
    \log \left(\frac{M_{A}^{1/2} +\1 } 2 \right) -
    \log \left(\frac{M_{A}^{1/2} -\1} 2 \right)\right) \biggr] .\nonumber
\end{eqnarray}
The rank 
\begin{eqnarray}
{\rm rank} \left[\frac{d}{dt}\left.
M_{A}^{1/2} (t)\right|_{t=0}\right] &= &
{\rm rank} \left[\frac{d}{dt}\left.
M_{A} (t)\right|_{t=0}\right]\nonumber\\
&\leq & 3 \, \text{rank}(h)
\end{eqnarray}
is upper bounded by an absolute constant, 
as is seen by explicitly computing the 
derivative in time. What is more, 
we find the upper bound
\begin{eqnarray}
\left\|
\left[\frac{d}{dt}\left.
M_{A}^{1/2} (t)\right|_{t=0}\right]
\right
 \| &=&
\| M_{A} (0)\| ^{-1/2} \nonumber
 \left\|
\left[\frac{d}{dt}\left.
M_{A} (t)\right|_{t=0}\right]
\right
 \| \\
 &\leq&
  \| M_{A} (0)\| ^{-1/2} 
  \frac{d}{dt}\|\gamma_A(t)\|  |_{t=0},
\end{eqnarray}
using the sub-multiplicativity of the operator norm several times.
Using the property that $\| M_{A} (0)\| \geq 1$,
this gives
\begin{equation}
\left\|
\left[\frac{d}{dt}\left.
M_{A}^{1/2} (t)\right|_{t=0}\right]
\right \|  \leq 
  \frac{d}{dt}\|\gamma_A(t)\|  |_{t=0}.
 \end{equation}
 Introducing the projection $\pi:= \1_A \oplus 0_B$,
 from this, we can upper bound the operator norm of the derivative and 
 conclude that 
\begin{eqnarray}
\left\|
\left[\frac{d}{dt}\left.
M_{A}^{1/2} (t)\right|_{t=0}\right]
\right
 \| &\leq&  
\| \pi ( \sigma h \gamma -\gamma \sigma h)\pi  \| 
\\
&\leq&  
\| ( \sigma h \gamma(0) -\gamma (0)\sigma h)\pi  \|  \nonumber
\\
&\leq& 
2  \|\gamma(0)\| 
  \|h\|  ,\nonumber
 \end{eqnarray}
 again using the sub-multiplicativity of the operator norm.
Then, the second term above can be bounded from above as
\begin{eqnarray}
\left \| \log \left(\frac{M_{A}^{1/2} \pm \1 } 2 \right)\right\|   &=&
\log \left(\frac{\left \|  M_{A} \right\| ^{1/2}\pm1 } 2 \right) \nonumber\\
&\leq& \log \left(\frac{\left \| \gamma(0) \right\| +1 } 2 \right).
\end{eqnarray} 
Putting these results together lets us arrive at the claim of the theorem,
giving rise to the monotone increasing function
$f:[1,\infty)\rightarrow \R$ that lets the small incremental entangling
bound  depends only on the coupling strength $\|h\|$.
\end{proof}

%
%

\bibliographystyle{apsrev}

\end{document}